\documentclass[twocolumn,prb,reprint,superscriptaddress,floatfix]{revtex4}
\usepackage[T1]{fontenc}
\usepackage[latin9]{inputenc}
\usepackage{amsmath}
\usepackage{graphicx}
\usepackage{amssymb}
\usepackage{hyperref}

\begin{document}

\title{Changing the state of a memristive system with white noise}

\author{Valeriy A. Slipko}
\email{slipko@univer.kharkov.ua} \affiliation{Department of Physics and Technology, V. N. Karazin
Kharkov National University, Kharkov 61077, Ukraine}
\author{Yuriy V. Pershin}
\email{pershin@physics.sc.edu} \affiliation{Department of Physics and Astronomy and University of South Carolina Nanocenter, University of South Carolina, Columbia, South Carolina 29208, USA}
\author{Massimiliano Di Ventra}
\email{diventra@physics.ucsd.edu} \affiliation{Department of Physics, University of California, San Diego, California 92093-0319, USA}

\begin{abstract}
Can we change the average state of a resistor by simply applying white noise? We show that the answer to this question is positive if the resistor has memory of its past dynamics (a memristive system). We also prove that,
if the memory arises only from the charge flowing through the resistor -- an ideal memristor -- then the
current flowing through such memristor can not charge a capacitor connected in series, and therefore cannot produce useful work. Moreover, the memristive system may skew the charge probability density on the capacitor, an effect
 which can be measured experimentally.
\end{abstract}

\maketitle

\section{Introduction}

If we connect a standard resistor to a random (white noise) voltage source, no average current flows in the system, and no change of resistance (state) of the
resistor can occur. This is simply because of the symmetry of the standard resistor with respect to positive and negative voltage fluctuations. However, there is now a renewed interest in a class of resistors with memory -- aptly called {\it memristors}~\cite{chua71a,chua76a} -- whose resistance varies according to the voltage applied to them, or the current that flows across them (for a recent review see, e.g., Ref. \onlinecite{pershin11a}). In this case, then, a fluctuation of the applied voltage may change the state of the memristor; and the ensemble of fluctuations could lead to a change of the {\it average} state of the memristor. If this is the case, what are the implications of the noise-induced state change? Is it possible, for example, to charge a capacitor through a noise-driven memristor to extract useful work?

In this paper we demonstrate analytically that the capacitor can not be charged through an {\it ideal} memristor (one whose state depends only on the charge flown through it) despite the change of the average state of such device.  Although we can not prove analytically a similar statement for the case of more general memristive systems,
our numerical simulations (for a particular device model and driving regime) also indicate the absence of capacitor charging. However, at least in the case of the ideal memristor,
we can monitor the change of its state by monitoring the {\it charge probability density} on the capacitor (which can be extracted by placing a voltmeter in parallel with the capacitor). This charge distribution probability density is {\it skewed} by the memory and could be detected experimentally. We focus here on an {\it external} noise source
because the thermal noise intrinsic to any resistor (and hence also to a memristor) cannot, by itself, be rectified~\cite{Brillouin50a}.

\begin{figure}[b]
\begin{center}
\includegraphics[width=.6\columnwidth]{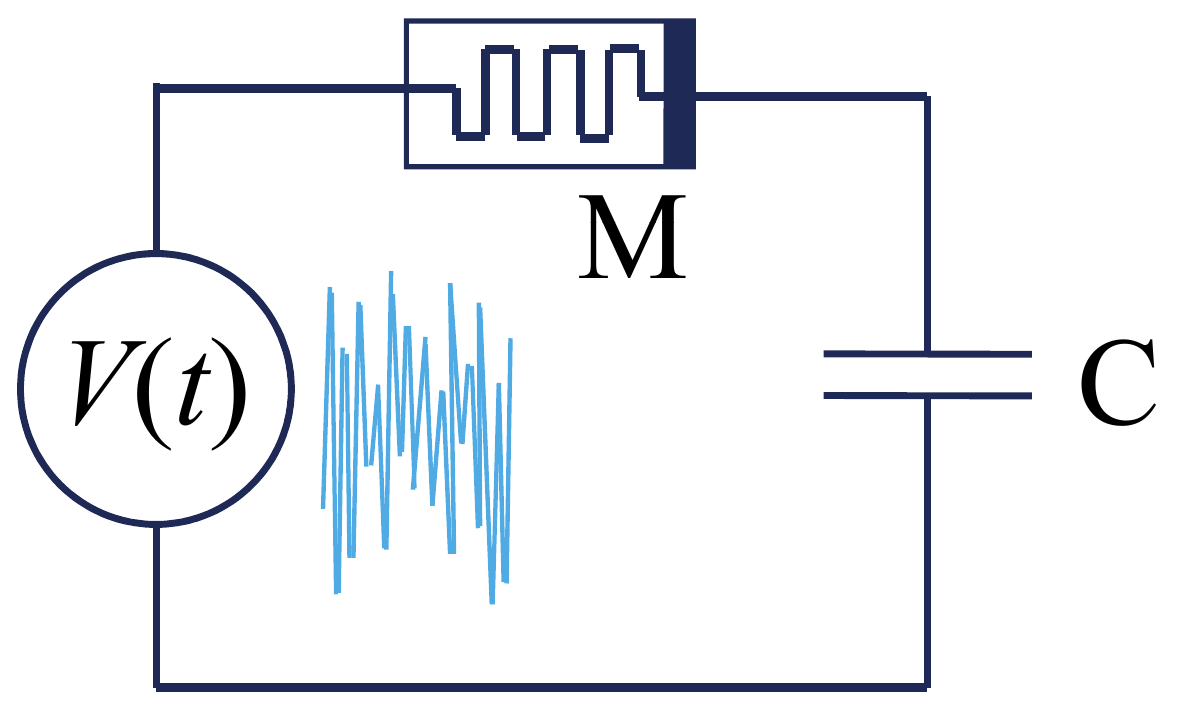}
\caption{\label{fig1} (color online). Circuit schematic: a stochastic voltage source $V(t)$ is connected to a memristive system M and capacitor C.}
\end{center}
\end{figure}

We note that memristive systems~\cite{chua76a} are particular types of circuit elements with memory~\cite{diventra09a,diventra09b}. There are two kinds of memristive systems: voltage-controlled and current-controlled ones~\cite{chua76a}.
The voltage-controlled memristive systems are defined by the equations
\begin{eqnarray}
I_M(t)&=&R^{-1}\left(x,V_M,t \right)V_M(t), \label{eq1a}\\
\dot{x}&=&f\left(x,V_M,t\right), \label{eq2a}
\end{eqnarray}
where $V_M(t)$ and $I_M(t)=\dot{q}(t)$ denote the voltage and current across the
device, $R$ is the memristance (memory resistance) and its inverse is the memductance (memory conductance), $x=\{x_i\}$ is a set of $n$ state variables describing
the internal state of the system, and $f$ is $n$-dimensional vector function. A current-controlled memristive system is such that
the resistance and the dynamics of state variables depend on the current \cite{chua76a,pershin11a}
\begin{eqnarray}
V_M(t)&=&R\left(x,I_M,t \right)I_M(t), \label{eq1-1}\\
\dot{x}&=&f\left(x,I_M,t\right). \label{eq1-2}
\end{eqnarray}
The ideal memristor that we consider below is a particular case of Eqs. (\ref{eq1-1}), (\ref{eq1-2}) when the memristance depends only on the charge flown through the device: $R=R(q)$. Memristive effects are not rare in nanostructures and can arise from different effects including ionic migration/redox reactions \cite{waser07a,yang08a}, spin polarization/magnetization dynamics \cite{pershin08a,wang09a}, phase transitions \cite{driscoll09b,wright2012phase}, etc. (see Ref. \onlinecite{pershin11a} for additional examples). 
The distinct feature of all memristive systems is the frequency-dependent pinched hysteresis loop \cite{chua71a,chua76a,pershin11a}. A previous study shows that the hysteresis of memristive elements can also be induced by white noise of appropriate intensity even at very low frequencies
of the external driving field \cite{Stotland12a}.

In this work, we consider the circuit shown in Fig. \ref{fig1} in which a memristive system M of memristance $R$ and a standard
capacitor C are connected to a Gaussian white noise voltage source $V(t)$. Our goal is to understand the circuit response and, in particular, to find the average values of memristance and capacitor charge.

The rest of this paper is organized as follows. In Sec. \ref{sec1} we consider the case of an ideal memristor and find analytically distributions and average values of the memristance and capacitor charge (Sec. \ref{sec11}). Then, we investigate the transient dynamics in the ideal memristor circuit (Sec. \ref{sec12}). Sec. \ref{sec2} presents a study of noise-driven voltage-controlled memristive system. Finally, in Sec. \ref{sec4} we give our conclusions.

\section{Circuits with Ideal Memristors} \label{sec1}

\subsection{Properties of steady state} \label{sec11}

We consider first the case of an
ideal memristor~\cite{chua71a}, whose memristance $R$ depends only on the
cumulative charge $q$ flown through the device. For the moment being, we do not select any specific form of $R(q)$ and only assume the existence
of a memory mechanism leading to an $R(q)$ dependence. For the circuit in Fig. \ref{fig1}, the equation of motion for $q$ is given by
\begin{equation}
R(q)\frac{\textnormal{d}q}{\textnormal{d}t}+\frac{q}{C}=V(t),
 \label{eq1}
\end{equation}
where $V(t)$ is a stochastic input signal. One can recognize that Eq. (\ref{eq1}) is a stochastic differential equation
of the Langevin type \cite{vanCampen}.  It is convenient to introduce a new variable $x$ instead of the charge $q$ as
\begin{equation}
x=\int\limits_{0}^{q}R(\tilde{q})\textnormal{d}\tilde{q}.
 \label{eq2}
\end{equation}
Since the memristance $R(q)$ is positive, $R(q)>0$, the dependence of
$x$ on $q$ given by Eq. (\ref{eq2}) is a one-to-one relation.
Consequently, Eq. (\ref{eq1}) can be rewritten in the form
\begin{equation}
\frac{\textnormal{d}x}{\textnormal{d}t}+\frac{q(x)}{C}=V(t).
 \label{eq3}
\end{equation}
For a given stochastic process  $V(t)$, Eq. (\ref{eq3}) determines the corresponding
stochastic process $x(t)$.
We assume that the stochastic
process $V(t)$ is Gaussian white noise,
\begin{equation}
\langle V(t)\rangle=0, \;\;\;\;\langle V(t)V(t^\prime)\rangle=2\kappa \delta(t-t^\prime),
 \label{eq4}
\end{equation}
where $\kappa$ is a positive constant characterizing the noise strength.

Instead of solving the nonlinear Langevin-type Eq. (\ref{eq3}), let us consider the corresponding Fokker-Planck equation (FPE)
\begin{equation}
\frac{\partial P(x,t)}{\partial t}=\frac{\partial }{\partial x}
\left(\frac{q(x)}{C}P(x,t)\right)+\kappa\frac{\partial^2 P(x,t)}{\partial x^2},
 \label{eq5}
\end{equation}
where $P(x,t)$ is the time-dependent charge probability density function. At this point, it is more convenient to return
to the initial variable $q$. We perform such a transformation taking into account
 the transformation law for the distribution function
\begin{equation}
P(x,t)=D(q,t)\frac{\textnormal{d}q}{\textnormal{d}x}=\frac{ D(q,t)}{R(q)},
 \label{eq6}
\end{equation}
 where $D(q,t)$ is the charge probability density function. Combining  Eqs. (\ref{eq5}) and  (\ref{eq6}) we find that the charge probability density function  $D(q,t)$ satisfies the following Fokker-Planck type equation
\begin{equation}
\frac{\partial D(q,t)}{\partial t}=\frac{\partial }{\partial q}
\left\{\frac{qD(q,t)}{CR(q)}+
\frac{\kappa}{R(q)}\frac{\partial }{\partial q}
\left(\frac{D(q,t)}{R(q)}\right)\right\}.
 \label{eq7}
\end{equation}

The FPE (\ref{eq7}) must be supplemented with an initial condition. For example, if at
$t=0$ the charge on the capacitor $q=q'$  with unit probability, then the initial condition for the charge probability density function
has the form $D(q,0)=\delta(q-q^{\prime})$, where $\delta(q)$ is the Dirac delta-function. In Sec. \ref{sec12} we will explicitly consider the transient dynamics of the probability density function, namely, the evolution of the initial condition into a stationary (equilibrium) solution of Eq. (\ref{eq7}).
Here, instead, we focus on the stationary solution $D_0(q)$ of FPE (\ref{eq7}) satisfying the following ordinary differential equation
\begin{figure}[tb]
\begin{center}
\includegraphics[width=.8\columnwidth]{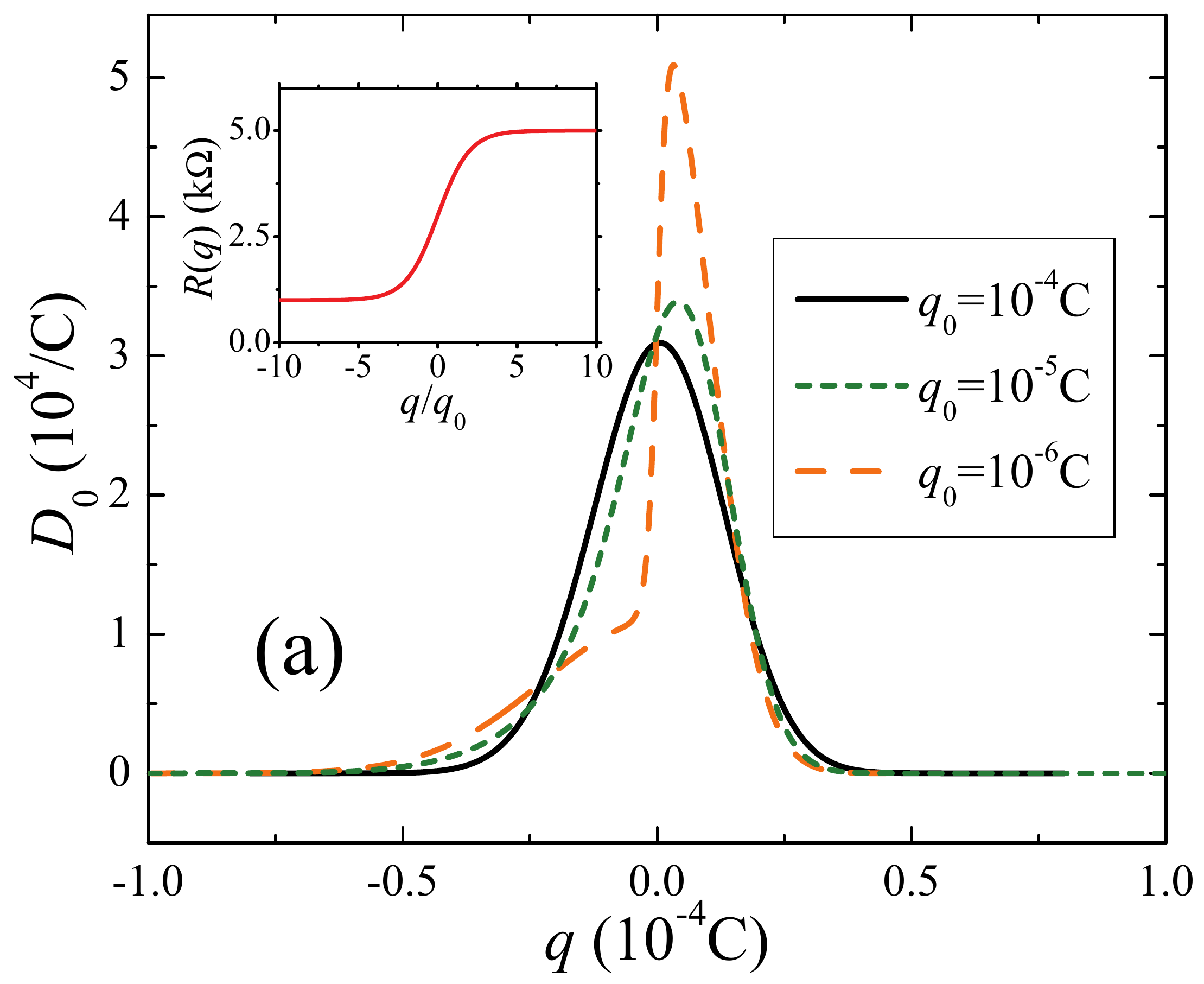}
\includegraphics[width=.9\columnwidth]{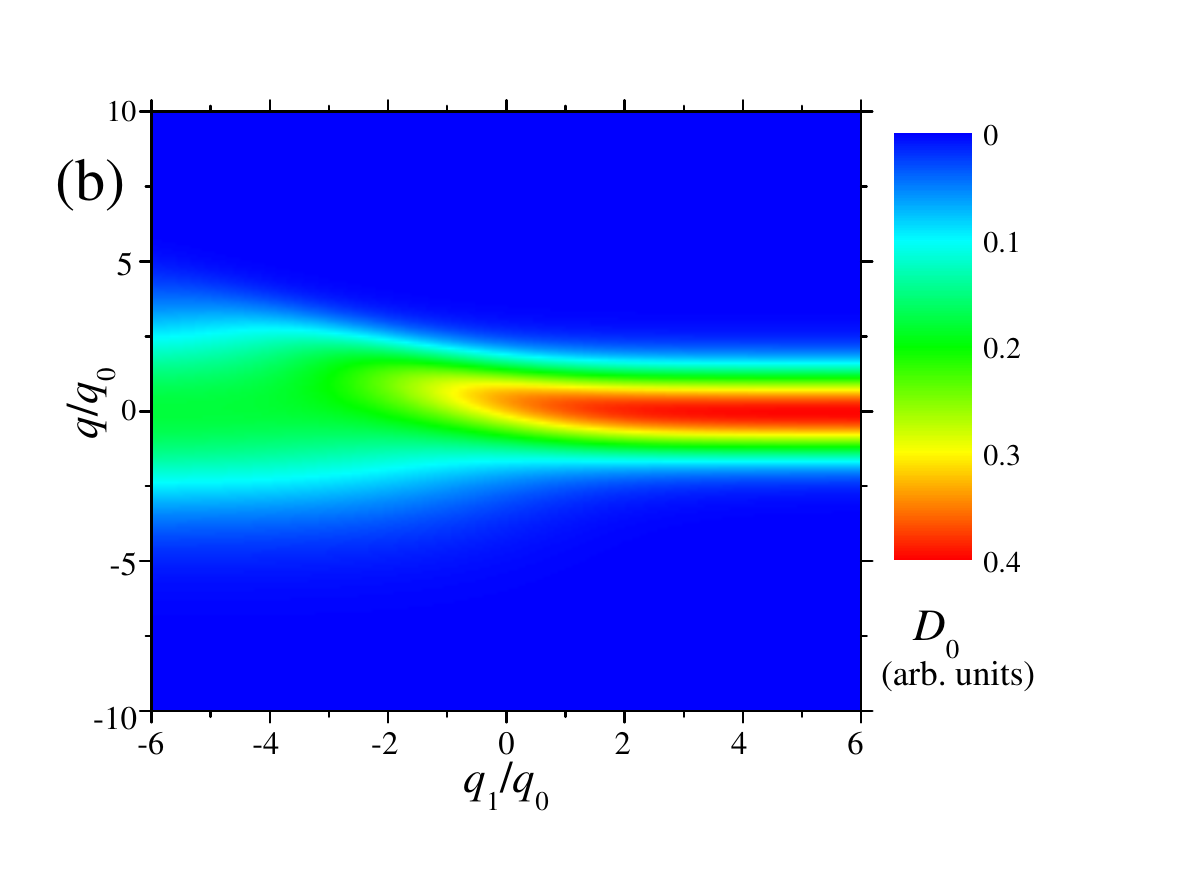}
\caption{\label{fig2} (color online). (a) Equilibrium charge probability density function $D_0(q)$ calculated assuming
$R(q)=R_{on}+\left( R_{off}-R_{on} \right)/(\textnormal{exp}\left[-(q+q_1)/q_0\right]+1)$ (shown in the inset).
This model describes a memristor whose memristance $R$ changes between two limiting values, $R_{on}$ and $R_{off}$.
The steepness of the transition between $R_{on}$ and $R_{off}$ is specified by a parameter $q_0$ which is a characteristic
charge required to switch the memristor. The constant $q_1$ is a parameter determining the memristance at the initial moment of time $t=0$.
The plot is obtained using  $R_{on}=1\textnormal{k}\Omega$, $R_{off}=5\textnormal{k}\Omega$, $C=1\mu$F,
$q_1=0$, $\kappa=0.5$V$^2$s for several different values of $q_0$ as indicated. (b) Equilibrium charge probability density function as a function
of $q_1$ calculated using the same model and parameters as in (a) at $q_0=10^{-5}$C.}
\end{center}
\end{figure}
 \begin{equation}
\frac{qD_{0}(q)}{CR(q)}+
\frac{\kappa}{R(q)}\frac{\textnormal{d} }{\textnormal{d} q}\left(\frac{D_{0}(q)}{R(q)}\right)=\textnormal{const}.
 \label{eq9}
\end{equation}
On physical grounds, we can safely assume that the memristance $R(q)$ acquires a limiting value at large values of $q$.
Then, it is not difficult to show that the general solution of Eq. (\ref{eq9}) is properly normalized ($\int D_0(q)\textnormal{d}q=1$) only
if the constant on the r.h.s. of Eq. (\ref{eq9}) is zero. Hence, the unique stationary solution of FPE (\ref{eq7}) is given by the following expression
\begin{equation}
D_0(q)=NR(q)\exp\left\{-\frac{1}{\kappa C}\int\limits_{0}^{q}\tilde{q}R(\tilde{q})\textnormal{d}\tilde{q}\right\},
\label{eq10}
\end{equation}
where $N$ is a normalization constant. Eq. (\ref{eq10}) clearly shows that the charge probability density function
is Gaussian only if $R=\textnormal{const}$. Any $q$-dependence of $R$ breaks such a property resulting in a non-Gaussian distribution function.
Typically, in experiments, the memristance switches between two limiting values \cite{pershin11a}. It then follows from Eq. (\ref{eq10}) that the tails of the
probability distribution function $D_0(q)$ are Gaussian
 \begin{equation}
D_0(q)\sim\exp\left\{-\frac{R(\pm\infty)q^{2}}{2\kappa C}\right\},~q\rightarrow\pm\infty,
 \label{eq10_1}
\end{equation}
 but asymmetric, since, normally, $R(-\infty)\neq R(+\infty)$.

 Fig. \ref{fig2}(a) presents the charge probability density function $D_0(q)$ calculated using Eq. (\ref{eq10}) with a specific form
 of memristance $R(q)$ specified in Fig. \ref{fig2} caption. When the parameter
$q_0$ is large (in this limit, the memristor approaches the behavior of a usual resistor since a larger charge is needed to change its state),
the charge probability density function is close to a Gaussian (solid line in Fig. \ref{fig2}(a)). Clearly, the probability density function gains an asymmetry with a decrease of $q_0$ (dashed lines in Fig. \ref{fig2}(a)). We note that under certain conditions a second maximum in the probability density function may develop. An example of such situation is shown in Fig. \ref{fig4}(c) below.

Moreover, it is important to emphasize that the charge probability density function $D_0(q)$ also depends on the initial state of the memristor
(defined by the parameter $q_1$ of the memristor model). Fig. \ref{fig2}(b) shows such a dependence for a selected set of parameters. The asymmetry in the charge probability density function is pronounced in
$|q_1/q_0| \lesssim  1$ and disappears when $|q_1/q_0|$ increases. The shift of $|q_1/q_0|$ from the region around 0 moves "the operational point" of the memristor into the saturation region where it behaves as a regular resistor.

 It is interesting to note  that, despite the asymmetry in the charge probability density function, the average value of the charge on the capacitor
 $\langle q\rangle_0=\int_{-\infty}^{+\infty} qD_0(q)dq$ in the stationary state $D_0(q)$ is {\it always} zero, as it follows from Eq. (\ref{eq10}):
\begin{eqnarray}
\langle q\rangle_0=-N\kappa C\int_{-\infty}^{+\infty}
 d\left[\exp\left\{-\frac{\int_{0}^{q}d\tilde q\tilde qR(\tilde
 q)}{\kappa C}\right\}\right]=0. \;\;\;
\label{eq10_1a}
\end{eqnarray}
This general result is the straightforward consequence of the mathematical structure of Eq. (\ref{eq10}).
Importantly, the property $\langle q \rangle_0=0$ does not depend on the specific form of $R(q)$.

\begin{figure}[tb]
\begin{center}
\includegraphics[width=.8\columnwidth]{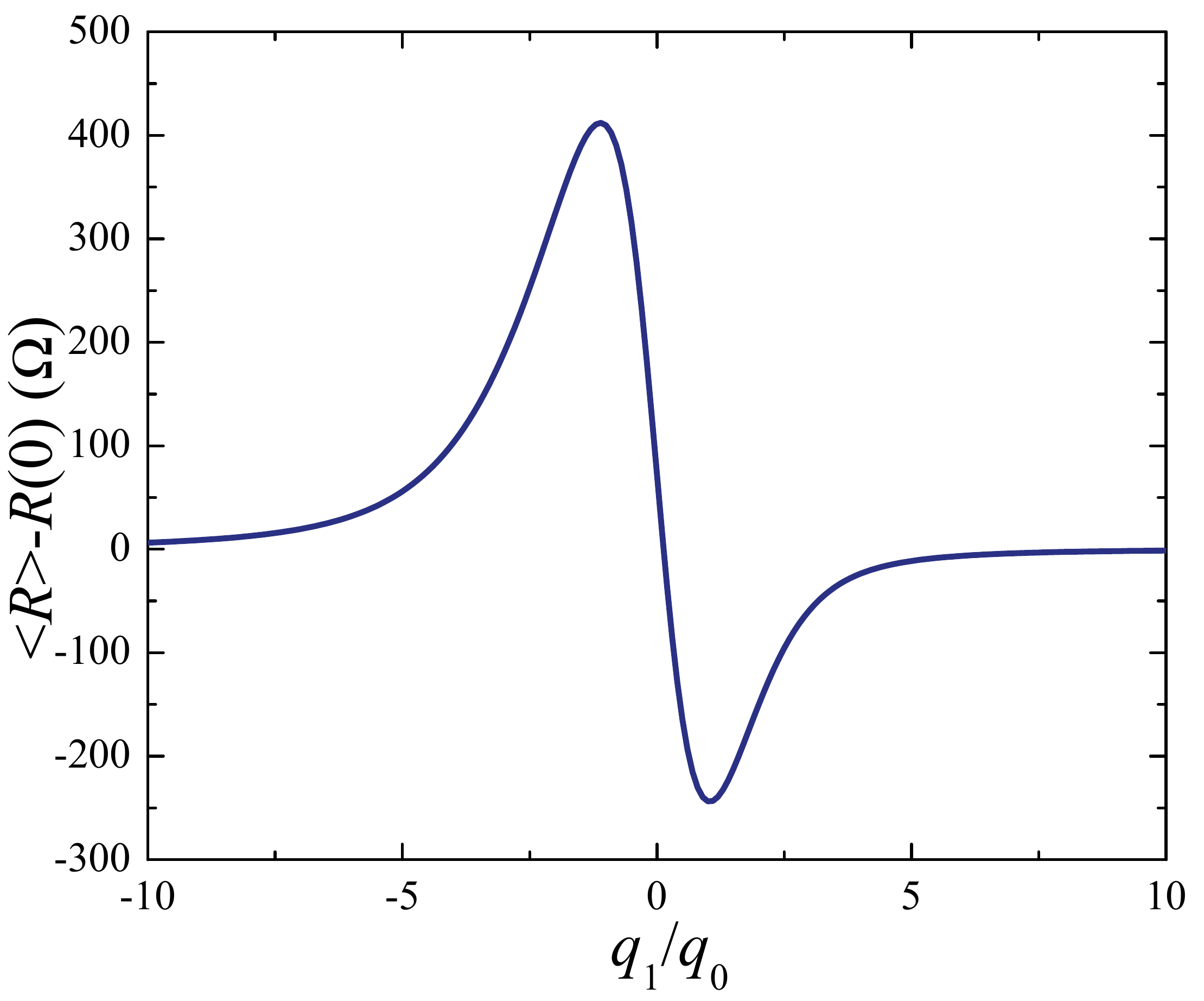}
\caption{\label{fig_x} (color online). Shift of the average value of memristance as a function of parameter $q_1$ specifying the initial value of memristance $R(0)$ in the following memristor model: $R(q)=R_{on}+(R_{off}-R_{on})\left( \textnormal{arctan} \left[ (q+q_1)/q_0\right]/\pi+0.5 \right)$. This plot is obtained for $R_{on}=1$k$\Omega$, $R_{on}=5$k$\Omega$, $q_0=10^{-5}$C, $C=1\mu$F and $\kappa=0.5$V$^2$/s.}
\end{center}
\end{figure}

However, a similar property does not hold for the average value of memristance $\langle R(q) \rangle$, which may be shifted from its initial value.
Since in the general case $R(q)$ is not linear in $q$, it is evident that $\langle R(q) \rangle =\int R(q) D_0(q)\textnormal{d}q \neq R(0)$. An example of such situation is shown in Fig. \ref{fig_x}, in which the initial state of the memristor is parameterized by a parameter $q_1$. Referring to Fig. \ref{fig_x}, the shift of the average value of memristance is mainly positive at negative values of $q_1$, and negative when $q_1$ is positive.

\subsection{Transient Dynamics} \label{sec12}

Next, we use the method of  separation of variables to find the general
time-dependent solution of Eq. (\ref{eq7}), which describes the transient
processes in the system. For this purpose, we select the specific solutions of Eq. (\ref{eq7}) in the form
 \begin{equation}
D^{(sp)}(q,t)=T_n(t)D_0(q)y_n(q).
 \label{eq10_2}
\end{equation}
Substituting Eq. (\ref{eq10_2}) into Eq. (\ref{eq7}) and separating
the variables, we obtain the following ordinary  differential equation for the unknown function
$y_n(q)$:
 \begin{equation}
-\frac{d}{dq}\left(\frac{\kappa D_0(q)}{R^2(q)}\frac{d y_n(q)}{dq}\right)=
\lambda_n D_0(q)y_n(q),
 \label{eq12}
\end{equation}
 where $\lambda_n$ are separation constants.  In order to be normalizable,
the specific solutions (\ref{eq10_2}) of Eq. (\ref{eq7})  must turn to zero
 at large $q$.  Thus the functions  $y_n(q)$ at infinity, $q\rightarrow\infty$,
 can grow, but not too rapidly.  This serves as the boundary condition for
the solutions $y_n(q)$ of Eq. (\ref{eq12}). In particular due to the asymptotic behavior (\ref{eq10_1}), the solutions  $y_n(q)$ can grow as a
power law at large $q$.

 The equation for functions $T_n(q)$ is trivially integrated, and it gives
the following solutions
\begin{equation}
T_n(t)=a_n e^{-\lambda_n t},
 \label{eq11}
\end{equation}
where $a_n$ are arbitrary constants.


 The general time-dependent solution of the Fokker-Planck equation (\ref{eq7}) can be presented
 as a sum of the specific solutions  (\ref{eq10_2})
with Eq.  (\ref{eq11}) taken into account
\begin{equation}
 D(q,t)=D_0(q)\sum^{+\infty}_{n=0} a_n e^{-\lambda_n t} y_n(q).
 \label{eq12_1}
\end{equation}
Constants $a_n$ can be determined from the initial condition for the probability density
function $D(q,0)$ by using the
weighted orthogonality of the solutions  $y_n(q)$ of Eq.  (\ref{eq12}),
\begin{equation}
\int_{-\infty}^{\infty} dq D_0(q)y_n(q)y_m(q)=0, ~~n\neq m,
 \label{eq12_1a}
\end{equation}
 which follows from the fact that Eq.  (\ref{eq12}) has the self-adjoint form.
 As a result we find
 \begin{equation}
 D(q,t)=\int_{-\infty}^{\infty} dq^\prime G(q,q^\prime,t)D(q^\prime,0),
 \label{eq12_2}
\end{equation}
 where
 \begin{equation}
 G(q,q^\prime,t)=D_0(q)\sum^{+\infty}_{n=0}\frac{e^{-\lambda_n t} y_n(q) y_n(q^\prime)}{\int_{-\infty}^{+\infty}dq D_0(q)y^2_n(q)}
 \label{eq12_3}
\end{equation}
 is the Green function of
FPE (\ref{eq7}), which corresponds to the initial condition $G(q,q^{\prime},0)=\delta(q-q^{\prime})$.

The value   $\lambda_0=0$ corresponds to the unique stationary state (\ref{eq10}), and from Eq. (\ref{eq10_2}) we conclude that $y_0(q)=1$.

It is impossible in general to integrate analytically Eq. (\ref{eq12}) for
$n\geq1$, or
 even to find the relaxation rate $\lambda_1$, which is the minimal nonzero
relaxation rate.  But we can make use of the variational approach to obtain a reasonable approximation for this rate.

Let us determine the functional acting on an arbitrary function $y(q)$
 as
 \begin{equation}
 F[y(q)]=
 \frac{\int_{-\infty}^{+\infty}dq\frac{\kappa D_0(q)}{R^2(q)}
 \left(\frac{dy(q)}{dq}\right)^2}{\int_{-\infty}^{+\infty}dq D_0(q)y^2(q)}
 \label{eq13}.
\end{equation}

 It is easy to show by using  integration by parts in the numerator of Eq.
(\ref{eq13}),  that the value of this functional
for the solution $y_n(q)$ of Eq.(\ref{eq12}) coincides with $\lambda_n$
 \begin{equation}
 F[y_n(q)]=-
 \frac{\int_{-\infty}^{+\infty}dq\frac{d}{dq}\left(\frac{\kappa D_0(q)}{R^2(q)}
 \frac{dy_{n}(q)}{dq}\right)y_n(q)}{\int_{-\infty}^{+\infty}dq D_0(q)y_{n}^2(q)}=\lambda_n.
 \label{eq14}
\end{equation}
Moreover, since the first variation of $F$ turns to zero  for the solutions of Eq. (\ref{eq12}), they are the stationary functions of functional (\ref{eq13}).

Note that because of the non-negativity of the functional, $F[y(q)]\geqslant 0$, we get
the same inequality for the relaxation rates $\lambda_n \geqslant 0$.

Straightforward calculation from Eq. (\ref{eq14})
shows that for an arbitrary function
$y(q)=\sum^{+\infty}_{n=0}c_n y_n(q)$ we find
\begin{equation}
 F\left[y(q)\right]=
 \frac{\sum^{+\infty}_{n=0}\lambda_nc_n^2}{\sum^{+\infty}_{n=0}c_n^2}
 \label{eq15}
\end{equation}

If $c_0=0$, i.e. a test function $y(q)$ is orthogonal to the function $y_0(q)=1$
in the sense that
 \begin{equation}
 \int_{-\infty}^{\infty}dqD_0(q)y(q)=0,
 \label{eq16}
\end{equation}
 then from Eq. (\ref{eq15}) it follows that for such functions
\begin{equation}
 F[y(q)]=
\frac{\lambda_1 c_1^2+\lambda_2 c_2^2+...}{c_1^2+c_2^2+...}
\geqslant\lambda_1.
 \label{eq17}
\end{equation}

Noting that the function $y(q)=q$ satisfies Eq. (\ref{eq16}) we  find the
following estimation for the relaxation rate
\begin{equation}
 F[q]=\kappa
 \frac{\int_{-\infty}^{+\infty}dq\frac{ D_0(q)}{R^2(q)}}
 {\int_{-\infty}^{+\infty}dq D_0(q)q^2}\geqslant\lambda_1.
 \label{eq18}
\end{equation}

 Thus the characteristic relaxation time $\tau$ of the system under consideration can be presented as a quotient of averages over the stationary state $D_0(q)$
 \begin{equation}
 \tau=\frac{\langle q^2\rangle_0}{\kappa\langle R^{-2}(q)\rangle_0}.
 \label{eq19}
\end{equation}

 When the resistivity $R(q)=R_0=const$,
i.e., if  we consider an ideal resistor, then Eq. (\ref{eq12}) becomes the Hermite differential equation. In this case we have
\begin{equation}
 y_n(q)=H_n\left( q\sqrt{\frac{R_0}{2\kappa C}}\right), ~\lambda_n=\frac{n}{CR_0},
 ~n=0,1, ...,
 \label{eq20}
\end{equation}
 where $H_n(x)$ is a Hermite polynomial.
 The stationary solution (\ref{eq10}) is the Gaussian distribution
 \begin{equation}
D_0(q)=\sqrt{\frac{R_0}{2\pi\kappa C}}\exp\left\{-\frac{R_0q^2}{2\kappa C}\right\},
\label{eq21}
\end{equation}
with $\langle q^2\rangle_0=\kappa C/R_0$,  and from Eq. (\ref{eq19})
 we find the well-known relaxation time of a RC circuit, $\tau=R_0C$.
Thus we see that for this case the estimate (\ref{eq18})
gives the exact
relaxation rate $\lambda_1=1/\tau=1/(R_0C)$. Note that in the case of constant
resistivity even the Green function (\ref{eq12_3}) can be calculated in a closed
form, being a Gaussian distribution with respect
to $q$ and $q^{\prime}$ at any moment of time t.

A better understanding of FPE solutions can be gained by noticing that
Eq. (\ref{eq7}) is similar to the drift-diffusion equation. Rewriting the right-hand side of
Eq. (\ref{eq7}) as
\begin{equation}
-\frac{\partial }{\partial q}
\left\{ \left[-\frac{q}{CR(q)}+\frac{\kappa}{R^3(q)}\frac{\textnormal{d}R(q) }{\textnormal{d} q} \right]D(q,t)-
\frac{\kappa}{R^2(q)}\frac{\partial D(q,t) }{\partial q} \right\},
 \label{eq77}
\end{equation}
we readily interpret the first term in Eq. (\ref{eq77}) as the drift and the second term as the diffusion term. Moreover,
the expression in the square brackets in Eq. (\ref{eq77}) plays the role of $e\mu E$ in the usual drift-diffusion equation, where $\mu$ is the mobility. Assuming $\mu=\textnormal{const}$, we introduce an effective electric field acting on the probability density
function as $E_{eff}=A[...]$ with $A$ is a positive proportionality constant and $[...]$ is from Eq. (\ref{eq77}). In the most simple situation, when $R=\textnormal{const}$, $E_{eff}=-Aq/(CR)$.
Notice that in this simple case $E_{eff}$ changes its sign at $q=0$ thus pushing the charge probability density function
toward the stable point $q=0$ from both positive and negative values of $q$.  The diffusion  term in Eq. (\ref{eq77}) tends to increase the distribution width. A balance between drift and diffusion is responsible for a finite distribution width.

\begin{figure}[tb]
\begin{center}
\includegraphics[width=.8\columnwidth]{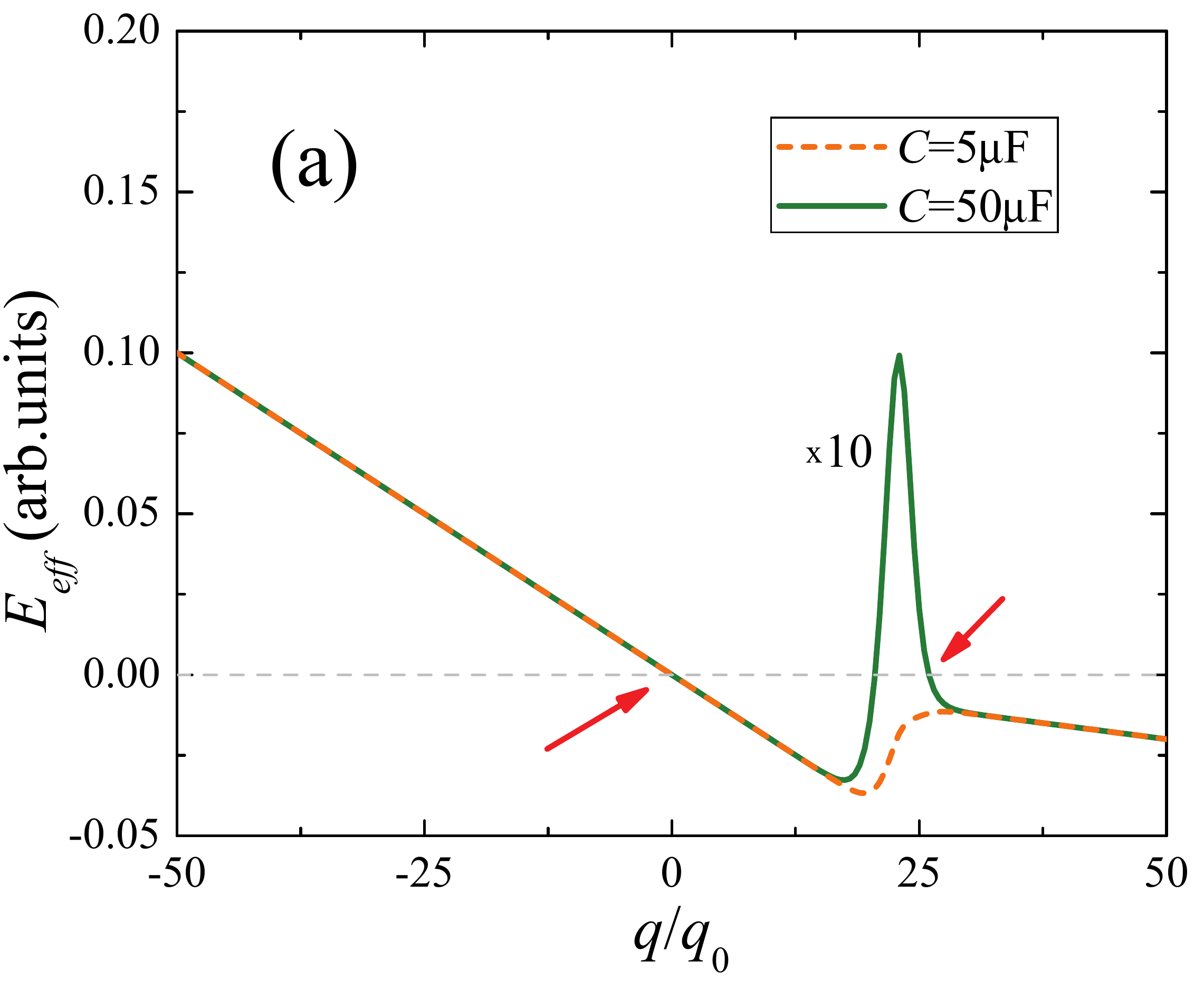}
\includegraphics[width=.8\columnwidth]{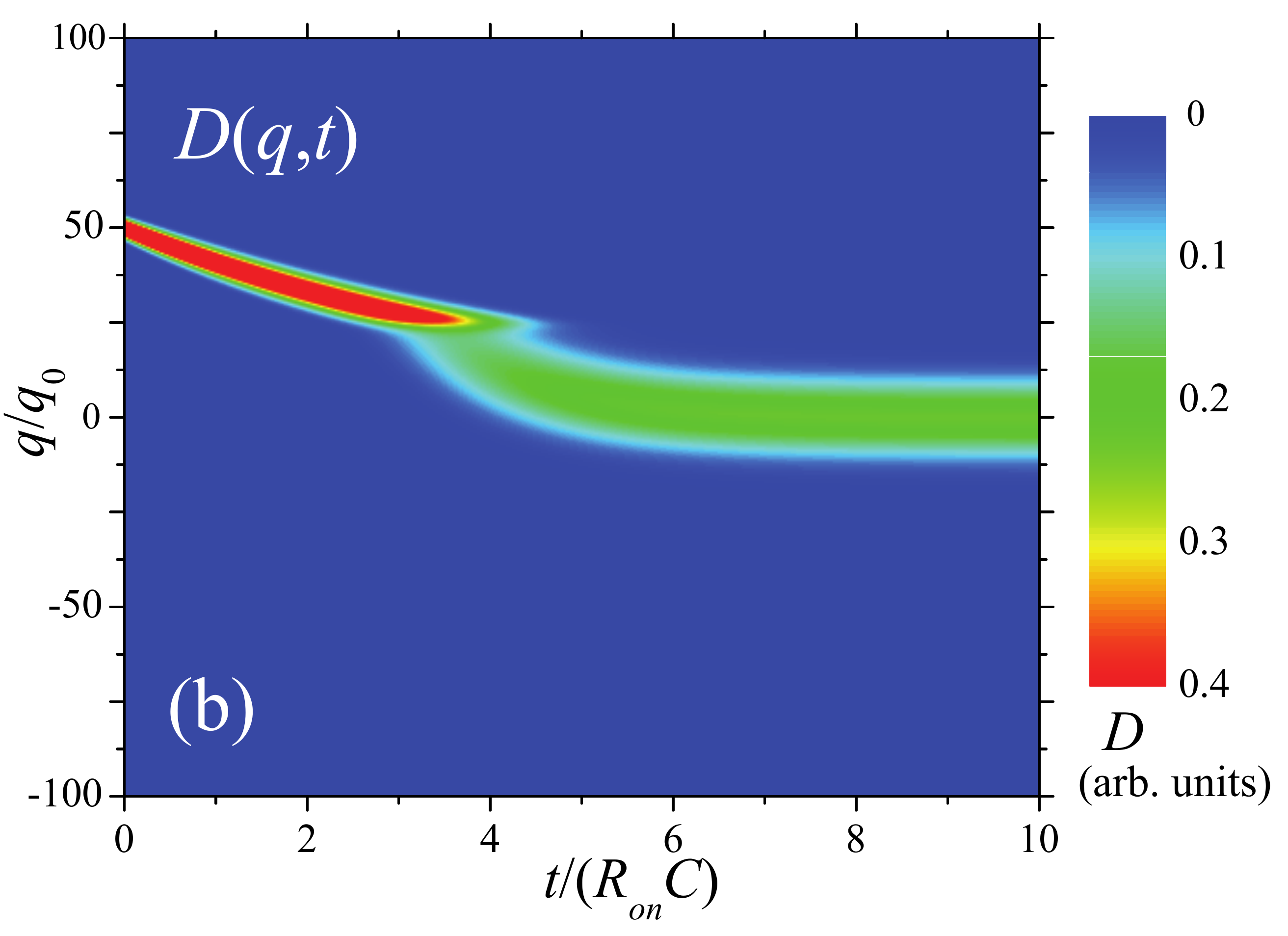}
\includegraphics[width=.8\columnwidth]{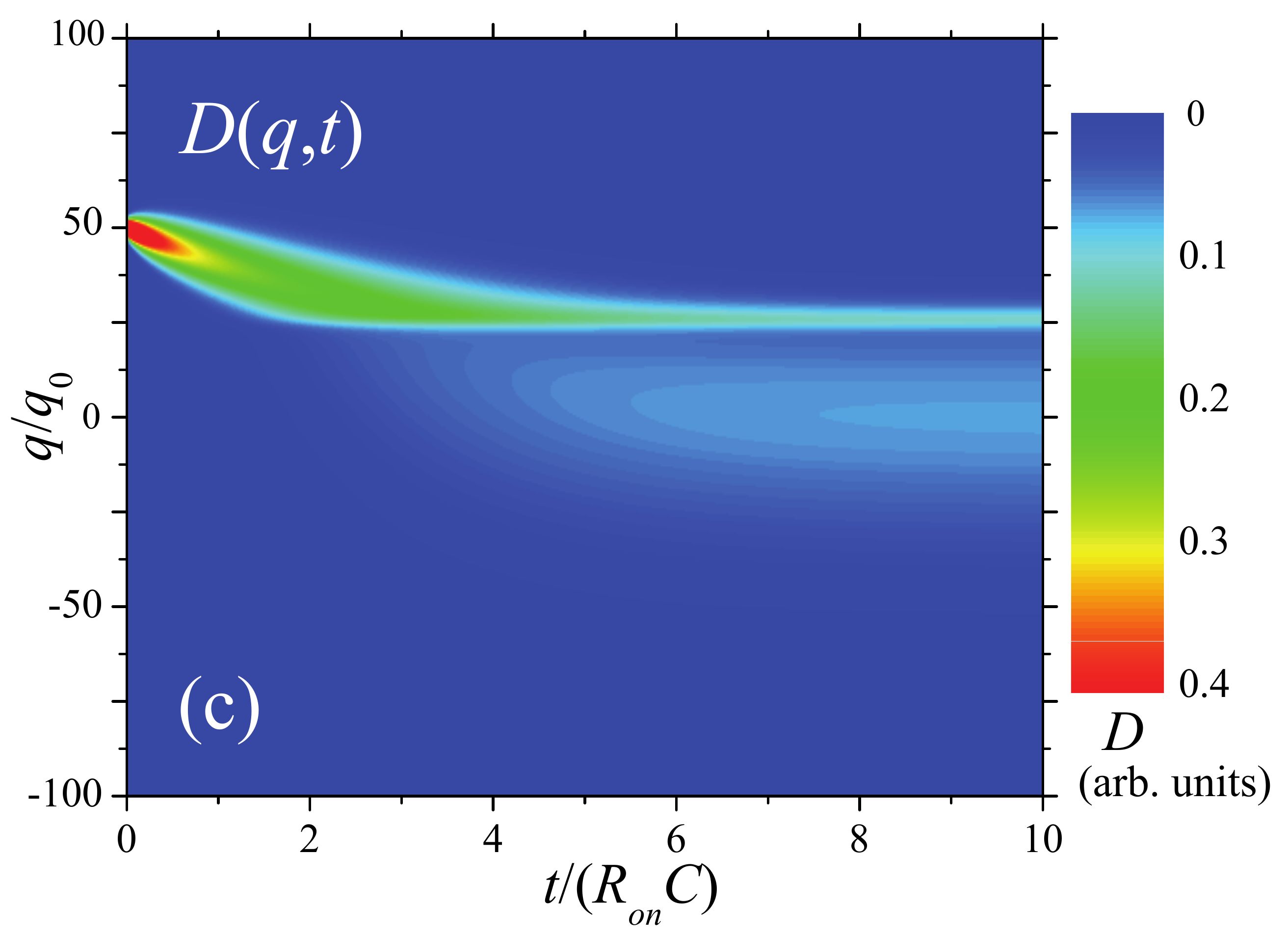}
\caption{\label{fig4} (color online). (a) An effective field $E_{eff}$ calculated for two different values of $C$ and memristor model specified in Fig. \ref{fig2} caption. Two stable points are denoted by arrows. The horizonal dashed line is for the eye. The calculation parameters are  $R_{on}=1\textnormal{k}\Omega$, $R_{off}=5\textnormal{k}\Omega$, $q_0=10^{-5}$C, $q_1=-0.00025$C, and $\kappa=1$V$^2$s.  (b) Dynamics of the charge probability density function at $C=5\mu$F. (c) Dynamics of the charge probability density function at $C=50\mu$F. (b) and (c) have been obtained assuming that the initial capacitor charge is narrowly distributed around $5\times 10^{-4}$C.}
\end{center}
\end{figure}

In the case of memristor, the expression for $E_{eff}$ acquires an additional contribution (the second term in the square brackets in Eq. \ref{eq77}).
Assuming that $R(q)$ is a monotonically increasing bounded function (e.g., as in Fig. \ref{fig2} caption model), this contribution can only locally increase
$E_{eff}$ in the region of $R(q)$ gradient. In certain cases such an increase has interesting consequences. Specifically, it may result in the development of additional  stable points as Fig. \ref{fig4}(a) exemplifies.

Figs. \ref{fig4}(b)-(c) present the dynamics of the charge probability density function for the case of one and two stable points (these plots correspond to the $E_{eff}$ curves in Fig. \ref{fig4}(a)). In the case of Fig. \ref{fig4}(b), the initially slower drift of $D(q,t)$ peak accelerates as the memristor passes through its switching region. At this point, the charge probability density function widens and then narrows back concentrating about $q=0$. The presence of two stable points
in the system results in a two-peak shape of the charge probability density function at longer times. Note, however, that  $<q>= 0$ at $t \rightarrow \infty$ as it follows from Eq. (\ref{eq10}).

\section{Circuits with Memristive systems} \label{sec2}

In this Section we consider the circuit shown in Fig. \ref{fig1} where M is a threshold-type memristive system. Such a configuration is of great interest
since many experimentally demonstrated memristive systems exhibit a threshold in their switching dynamics (see, for example, Refs. \onlinecite{tao2006electron,waser07a,Sawa08a,pershin11a}). However, mathematical/computational modeling of such cases in the presence of noise is complicated by non-linear noise terms entering the equations of system dynamics. In fact, accurate mathematical/computational approaches to treat such situations still need to be developed. Here, we study the circuit dynamics based on some intuitive arguments complemented by numerical results found for a linearized model.

Let us consider a specific regime of circuit operation when fluctuations of the input voltage source
(Gaussian white noise is assumed, see Eq. (\ref{eq4})) are smaller then the threshold voltage of the memristive system $V_t$, so that the voltage across the memristive system M is smaller than $V_t$ for the most of the time. In this regime, the switching events of memristance are relatively rare. Their intensity is determined by the voltage fluctuations across M given by $V_M=V(t)-V_C$, so that fluctuations of both $V(t)$ and $V_C$ are important.

One can notice that during the intervals of constant $R$, the fluctuations of $V_C$ are described by the Ornstein-Uhlenbeck process. Consequently,
\begin{equation}
\textnormal{Var}\left[ V_C(t)\left|V_C(0)\right.=0\right]=\frac{\kappa}{RC}\left(1-e^{-\frac{2t}{RC}} \right), \label{eq:varVc}
\end{equation}
where $2 \kappa$ is the noise strength of $V(t)$. If we use Eq. (\ref{eq:varVc}) as an estimate for the amplitude of typical voltage fluctuations across the memristive system then it follows that such typical fluctuations are weaker for larger values of $R$ and stronger when $R$ is smaller. Consequently, we expect that the memristive system M spends less time in states with smaller $R$ (since the probability of switching from these states is higher due to stronger fluctuations) and more time in states with larger $R$. Our qualitative prediction, thus, is a rather larger value of $\left< R\right>$.

\begin{figure}[tb]
\begin{center}
\includegraphics[width=.8\columnwidth]{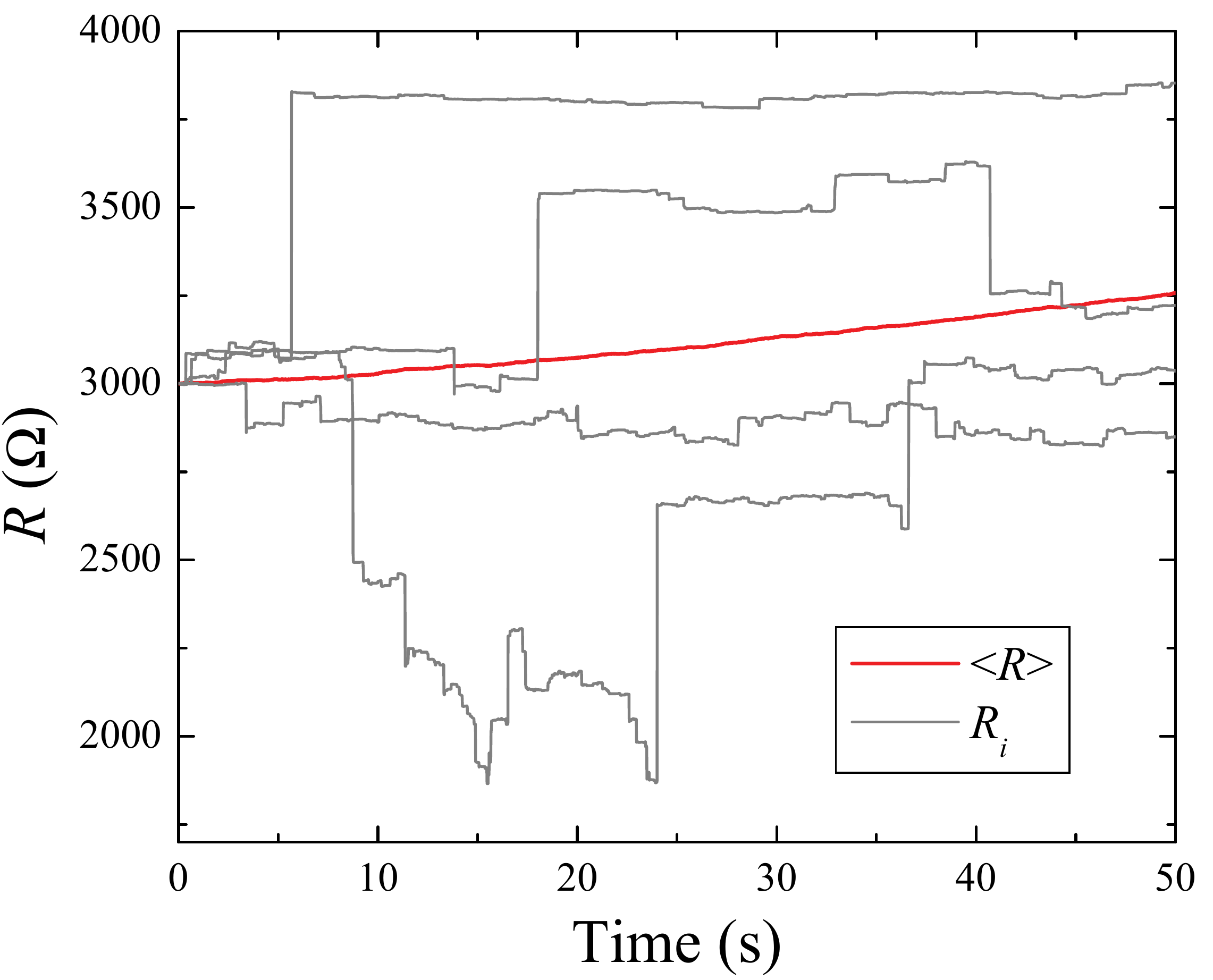}
\caption{\label{fig3} (color online). Simulations of the circuit shown in Fig. \ref{fig1} with a voltage-controlled memristive system (Eq. (\ref{sinh_model})). The curves show the memristance averaged over 5000 realizations ($\left< R\right>$) and several examples of particular realizations of memristance ($R_i$). This plot was obtained using the parameter values $\alpha=0.1\Omega$/s, $R_{on}=1$k$\Omega$, $R_{off}=5$k$\Omega$, $R(t=0)=3$k$\Omega$, $V_t=0.2$V, $\sqrt{2 \kappa}=0.1$V$\sqrt{\textnormal{s}}$.}
\end{center}
\end{figure}

In order to test this prediction, let us consider a specific model of a voltage-controlled memristive system with a ``soft'' threshold such that Eq. (\ref{eq2a}) is written as
\begin{equation}
\dot{x}=\alpha \;\;\textnormal{sinh}\left(\frac{V_M}{V_t}\right), \label{sinh_model}
\end{equation}
where $\alpha$ is a constant, $V_t$ is the threshold voltage and $x\equiv R$. It is also assumed that $R_{on}\leq R \leq R_{off}$. The circuit shown in Fig. \ref{fig1} is modeled by a couple of stochastic differential equations describing evolution of stochastic variables $q$ and $R$. We linearize Eq. (\ref{sinh_model}) with respect to small values of the input $V(t)$ and solve the two linear stochastic differential equations numerically \cite{mannella2002}. Some results of our simulations are presented in Fig. $\ref{fig3}$. This plot shows that the average value of memristance $R$ increases in time in agreement with the above discussion. Moreover, for the selected values of parameters, our numerical simulations do not reveal any significant deviations of the average voltage across the capacitor from zero, and asymmetry in the charge probability density function. We emphasize that our numerical results should be considered mainly qualitatively as the linearization procedure is valid only for small fluctuations of $V(t)$. At the same time, these results support our qualitative considerations above, e.g., regarding the average value of $R$.

The dependence of the variance of $V_C$ on $R$ given by Eq. (\ref{eq:varVc}) can also be applied to understand the asymmetry of the noise distribution function shown in Fig. \ref{fig2} for the case of ideal memristors. We recall that in these devices the memristance $R$ is a function of $q$ only, namely, $R=R(q)$. Consequently, when $q$ is positive and $R$ is large, the voltage fluctuations (according to Eq. (\ref{eq:varVc})) are reduced and, consequently,  the charge probability density function is narrower. In the opposite case of negative $q$, the voltage fluctuations are increased (because $R$ is smaller) and the charge probability density function is wider. This is exactly the same behavior as observed in Fig. \ref{fig2}. We anticipate that in the case of memristive systems a similar change of the charge probability density function is also possible in the regime of strong noise, when the memristive system stays under switching conditions during a significant fraction of the time evolution. However, in the case of weak noise considered here, the correlation between charge flown through the memristive system and its state is almost negligible, and therefore we do not expect any significant asymmetry of the charge probability density function in this regime.

\section{Conclusion} \label{sec4}

To summarize, we have shown that the charge probability density function may be modified in memristive circuits coupled to white noise sources.
In the specific circuit example that we have considered (memristor and capacitor driven by a stochastic voltage source),
the distribution gains an asymmetry that disappears if we replace the memristor by a usual resistor. We have proved analytically that for any
charge-controlled memristor, the average charge on the capacitor is zero. This is a very surprising result taking into account the fact that the
memristor introduces a circuit asymmetry. We have also developed a formalism to describe the evolution of the charge probability density function. It can be used to describe non-equilibrium processes in the circuit (for example, the discharge of
the initially charged capacitor). We finally note that our theoretical predictions can be easily tested
experimentally with available memristive systems.

\section*{Acknowledgment}

This work has been partially supported by NSF grants No. DMR-0802830 and ECCS-1202383, and the Center for Magnetic Recording Research at UCSD.

\bibliography{memcapacitor}

\end{document}